\documentclass[12pt,twoside,a4paper,english,openright,fleqn,onecolumn]{article}

\usepackage[english]{babel}
\usepackage{graphicx,amssymb,multicol}
\usepackage{epsfig} 
\usepackage[T1]{fontenc}
\usepackage{authblk}

\psfull

\title{\vspace{-1.1cm} \normalsize \textbf{\textsc{CF3/CF6 White Paper:\\ \Large The hunt for axionlike particles with the Cherenkov Telescope Array \vspace{-0.5cm}}}}
\author[1]{\normalsize M. A. S\'anchez-Conde\thanks{masc@stanford.edu}}
\author[1]{\normalsize S. Funk}
\author[2]{\normalsize F. Krennrich}
\author[2]{\normalsize A. Weinstein}
\affil[1]{\footnotesize \it KIPAC/SLAC National Accelerator Laboratory, Menlo Park, CA 94025, USA}
\affil[2]{\footnotesize \it Department of Physics and Astronomy, Iowa State University, Ames, IA 50011, USA}

\begin{document}
\date{}

\maketitle

\vspace{-1cm}

\section*{\normalsize \textbf{\textsc{Axionlike particles and the gamma-ray connection}}} \label{sec1}
\vspace{-0.2cm} 

At present, the Peccei-Quinn mechanism remains the most convincing solution for the strong Charge-Parity problem in Quantum Chromodynamics \cite{PQ}. As early as 1978, Weinberg \cite{weinberg} and Wilczek \cite{wilczek} realized independently that a consequence of this mechanism is the existence of a pseudo-scalar boson, the axion. One generic property of axions is a two-photon interaction of the form ${\cal L}_{a \gamma} = -\frac{1}{4M}~F_{\mu \nu}\overline{F}^{\mu \nu}a = \frac{1}{M}~{\bf E \cdot B}~a$, where $a$ is the axion field, $M$ is the inverse of the photon/axion coupling strength, $F$ is the electromagnetic field-strength tensor, $\overline{F}$ its dual, ${\bf E}$ the electric field, and ${\bf B}$ the magnetic field. The axion has the important feature that its mass $m_a$ and coupling constant are inversely related to each other. There are, however, other predicted states that exhibit the same phenomenology but for which this relation between mass and coupling does not hold; such states are known as Axion Like Particles (ALPs). An important and intriguing consequence of the previous Lagrangian is that axions convert into photons and vice-versa in the presence of an electric or magnetic field \cite{dicus,sikivie}. Indeed, this effect is the basis of ongoing axion searches in the laboratory \cite{cast,admx}. The conversions of ALPs may have important implications for astrophysics since they would imprint characteristic features on the spectra of gamma-ray sources, as we will discuss in more detail later and as it is discussed, e.g., in Refs.~\cite{hooper,deangelis,mirizzi07,mena,wouters}. The features may be detected by the {\it Fermi} Gamma-ray Space Telescope and by current Imaging Atmospheric Cherenkov Telescopes (IACTs) like MAGIC, H.E.S.S., VERITAS or the future Cherenkov Telescope Array (CTA). In addition, both axions and ALPs  are also appealing because they can constitute a significant fraction of the dark matter content of the Universe. This is the topic of the CF3 working group.\\

The gamma-ray spectra of distant astrophysical sources therefore can be a powerful probe of physics beyond the Standard Model. Indeed, some recent observations of distant Active Galactic Nuclei (AGNs) seem to be at odds with the standard astrophysical models that explain the emission mechanisms in the gamma-ray source combined with the state-of-the-art Extragalactic Background Light (EBL) models. (The EBL is the topic of the CF6 working group). In particular, unexpectedly high fluxes have been measured at the highest energies, where absorption by the EBL is most severe \cite{hess06,h&k12,horns12}. The hard spectra deduced for some AGNs are difficult to explain with conventional physics as well (see, e.g., Ref.~\cite{angelis_review}). Also, a few AGNs exhibit the so-called pile-up problem, i.e. a deviation from the expected power-law intrinsic spectra at the highest observed energies \cite{tigu11}. While these observations can be explained with conventional physics (e.g., particles in blazar jets accelerated at relativistic shocks~\cite{stecker07}, internal photon-photon absorption~\cite{aharonian08}, and/or secondary gamma rays produced along the line of sight by the interactions of cosmic-ray protons with background photons~\cite{essey10,essey11,aharonian12}), photon/ALP conversions offer a single viable explanation that can simultaneously resolve all these puzzles. (Lorentz Invariance Violation would represent another exciting possibility, as also discussed in the CF6 working group). CTA, with its excellent sensitivity \cite{CTA}, will be able to strongly constrain such a scenario for the first time in the TeV energy range, as detailed below.

\section*{\normalsize \textbf{\textsc{Current status of the search for ALPs in gamma rays}}} \label{sec2}
\vspace{-0.2cm} 

Photon/ALP conversion can happen in any astrophysical environment that provides the correct combination of magnetic field strength and size of the confinement region of the magnetic field. Intergalactic magnetic fields (IGMFs)  \cite{deangelis,mirizzi07,alps_Masc}, the Galactic magnetic field \cite{simet}, magnetic fields typical of galaxy clusters  \cite{horns12b} and those of gamma-ray sources such as AGNs \cite{hooper,tavecchio} have all been predicted to give rise to measurable effects. The ALP-induced spectral features vary from one scenario to another. Conversions taking place at the source (as with those caused by the magnetic fields of AGNs) primarily give rise to a flux attenuation, as some photons that escape from the source just convert into ALPs and therefore do not reach the Earth. Conversion in IGMFs can attenuate or enhance the flux depending on distance, magnetic field and, most importantly, on the intensity of the EBL. Although some photons convert into ALPs (which leads to flux attenuation), flux enhancement is still possible because when the photon is converted into an ALP it does not interact with the EBL, and thus is not absorbed. Later, a fraction of those ALPs convert back into photons before reaching the observer, potentially resulting in an enhanced flux. \\

Gamma-ray observations can provide valuable constraints on the nature of ALPs, and therefore represent an excellent complementary approach to laboratory searches for ALPs, in terms of those ALP masses and coupling constant values that can be tested (see, e.g.,~\cite{brun13} for a recent review). In the TeV energy range, the first effort aimed at analyzing IACT data to look for ALP spectral features \cite{horns12} involved a statistical analysis of a large sample of very high energy AGN spectra (50 in total) and found a significant anomaly (suppression) of the pair production process at a 4.2$\sigma$ level, i.e., that the correction for absorption with current EBL models is too strong for the measurements at the energies with the greatest attenuations. The measured effect is furthermore compatible with that expected from proposed photon/ALP scenarios. Very recently, based on this finding and results for a few more spectra of distant AGNs, the same authors set, for the first time, lower limits to the photon/ALP coupling constant by considering different possible configurations of astrophysical magnetic fields and different astrophysical environments, under the assumption that the tension between TeV measurements and EBL models is real \cite{meyer13}. These lower limits are found to be competitive with those obtained in the lab, and are of the order of 2$\times$10$^{-11}$ GeV$^{-1}$ in the conservative scenario they considered. This value is close to the latest upper bounds of CAST \cite{cast12} and within the sensitivity estimates of future experiments such as ALPS-II \cite{alpsII} and IAXO \cite{iaxo}. On the other hand, another approach has recently been used to set lower limits to the photon/ALP coupling constant:  irregularities in the AGN spectra induced by the turbulence of the IGMF can be predicted on a statistical basis and then used to constrain the ALP parameters~\cite{wouters12}.  The limits obtained in this way from H.E.S.S. observations of PKS 2155$-$304 improve the CAST limits in a limited ALP mass range around 20 neV \cite{wouters12}.

\section*{\normalsize \textbf{\textsc{The ALP search with CTA}}} \label{sec3}
\vspace{-0.2cm} 

\begin{figure} [!t]  
\centering
\begin{minipage}[c]{0.6\textwidth}
 \centering \includegraphics[width=9.8cm]{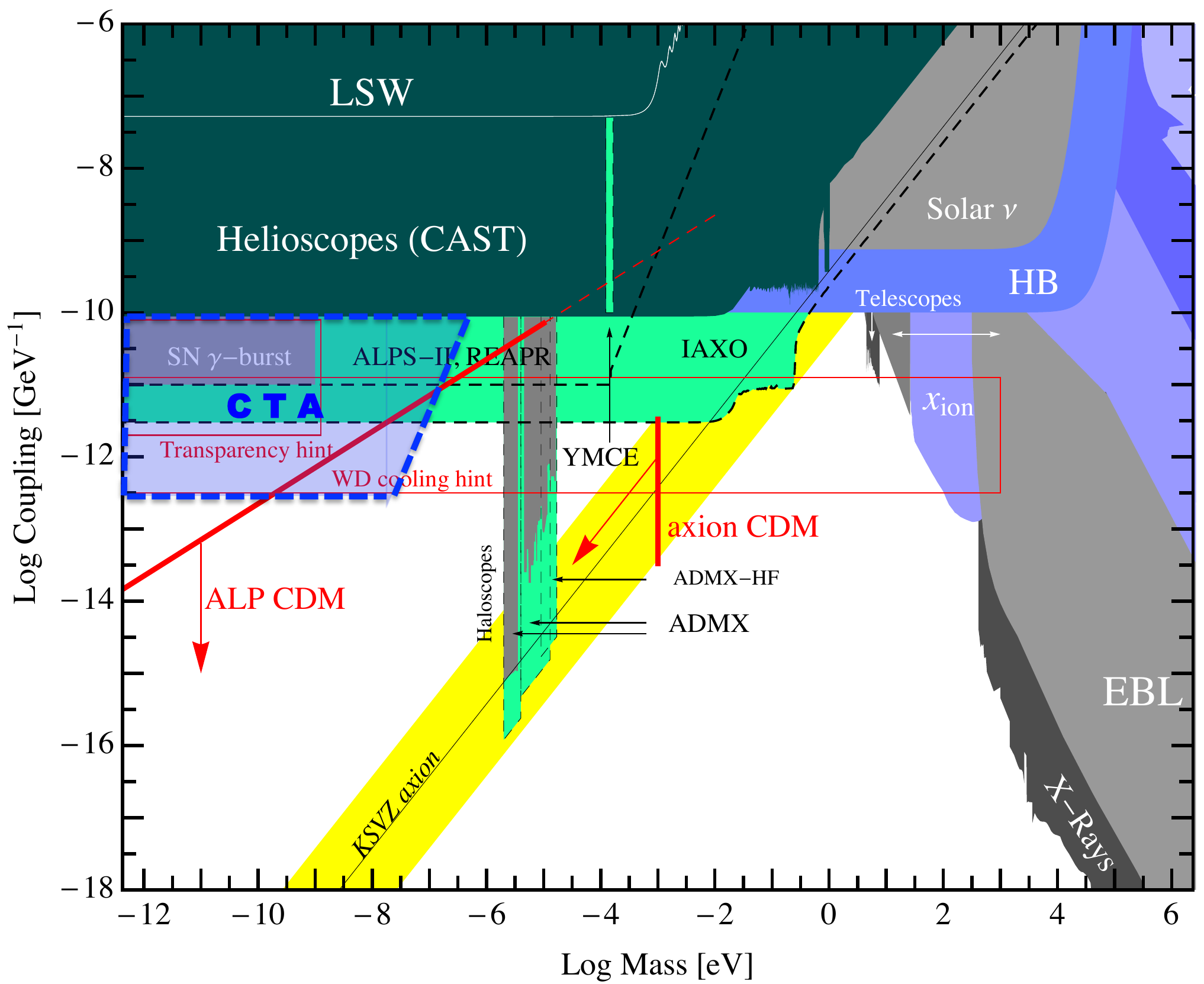}
\end{minipage}%
 \begin{minipage}[c]{0.4\textwidth}
\centering \includegraphics[width=6.8cm]{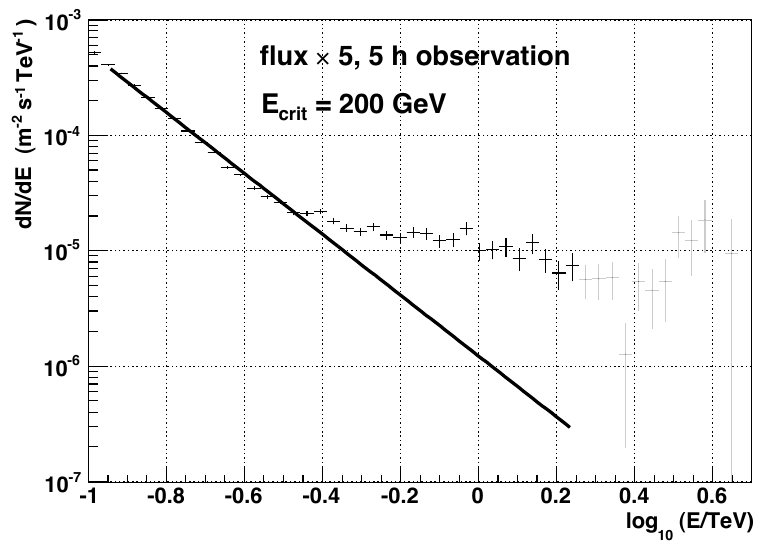} 
\end{minipage} 
\caption{\footnotesize{{\it Left panel:} An estimate of the ALP parameter space {\it accessible} to CTA. The CTA region (delimited by the dashed blue line) is difficult to explore with other detection techniques. Indeed, at present there is no experiment that can probe that region. ALPS-II \cite{alpsII} and IAXO \cite{iaxo} are still to come (light green indicates future prospects). The region labelled as "SN $\gamma$-burst" corresponds to the exclusion limits implied using the non-detection of gamma rays from the supernova 1987A during the 10-s time window defined by the neutrino burst \cite{brockway, griffols}. However these limits are subject to large uncertainties (see, e.g., section III.C in \cite{alps_Masc}). The red thin lines that can be seen in the CTA region are not actually experiments, but only hints coming from different astrophysical scenarios (issues with the rate of cooling in white dwarfs \cite{WD}, or lack of opacity in the Universe to $\gamma$-rays). Such hints do not exclude anything, and have been included here only to show the range of ALP parameters that might explain those astrophysical puzzles. Note also that, very interestingly, CTA could probe a region in which ALPs might behave as cold dark matter in a region inaccessible to other current or planned experiments. The figure should be taken with some caveats though. The excluded regions were derived from laboratory experiments (where the environment is well known) or from experiments using the Sun or stars (where the physical environment can be estimated with reasonable accuracy). As the search for ALPs with gamma-ray observations involves large uncertainties (EBL, IGMF...), it will not be trivial to constrain the ALP parameters if subtle spectral features are observed. Yet the figure is useful as it shows the region {\it plausibly reachable} by CTA. As discussed in the text, detailed simulations of photon/ALP conversion processes for different astrophysical environments, magnetic values, and ALP parameters will be needed in order to {\it confidently} set constraints on such a portion of the ALP parameter space. Adapted from \cite{ringwald}. {\it Right panel:} Simulation of a 5~h CTA observation of a flare of 4C 21.35 five times more intense than the one recorded by MAGIC \cite{pks1222_MAGIC}. The black points indicate the energy bins used for the fit, i.e. those with a signal exceeding three times the RMS of the background, and a minimum of 10 excess events (excluded points in grey). The estimated intrinsic differential energy spectrum (after correction with the Dom\'inguez et al. EBL model \cite{dominguez11}) is brightened at high energies due to photon/ALP conversions. The IGMF strength is assumed to be 0.1 nG in this example. Taken from \cite{CTAaxions}.}}
\vspace{-0.25cm}
\label{fig:axions}
\end{figure}

CTA's twenty times larger collection area and improved sensitivity compared to current IACTs \cite{CTA} will make it an extraordinarily powerful instrument to search for signatures of ALPs imprinted in the spectra of gamma-ray sources in the energy range between a few tens of GeV up to a few dozen TeV.  Very importantly, CTA will explore a range of ALP parameters that is otherwise very difficult to reach, as it can be seen in the left panel of Fig.~\ref{fig:axions}. Detailed simulations of photon/ALP conversion processes for different astrophysical environments, magnetic field strengths and confinement volumes, and ALP parameters will be needed in order to {\it confidently} set constraints on this part of the ALP parameter space. \\

Preliminary studies for CTA have already simulated CTA observations of AGNs in the presence of relevant photon/ALP mixings~\cite{CTAaxions}. More specifically, a source was simulated based on the flat spectrum radio quasar 4C +21.35 (PKS 1222+21, z = 0.432), assuming an intrinsic unbroken power-law spectrum, and different absolute flux normalizations and flare durations. The distortion of the spectra due to ALPs depends on the particular case, but generally the larger the intensity or the duration of the flare, the larger the accessible range of ALP parameters that could be tested by CTA. As an example, it was found that a 0.5 h duration flare like the one reported by MAGIC \cite{pks1222_MAGIC} would not be enough for CTA to detect a significant effect in any of the ALP scenarios tested, i.e., those described in Ref.~\cite{alps_Masc}. However, a flare of similar intensity, but lasting 5 h, would already be enough to detect the predicted ALP spectral features for the same ALP scenarios (see the right panel of Fig.~\ref{fig:axions}). Not only PKS 1222+21 but many other similar objects could be monitored by CTA (e.g., probably more than 100 above this same redshift~\cite{yoshiCTA}).\\

The most suitable energy for ALP searches with CTA is an intermediate one in which the effects of attenuation by the EBL are already present but introduce only a moderate absorption, i.e., from $\sim$100 GeV up to a few TeV. Thus, the US contribution of Medium Size Telescopes is particularly important  for ALP searches, as it will improve the sensitivity of the core energy regime (100 GeV - 10 TeV) of CTA by a factor of 2-3 \cite{jogler}. Outside this energy range, and even when ALP spectral features lie within the energy range covered by CTA, they might not be accessible in practice. At the highest energies, from several to tens of TeV, the attenuation due to the EBL for a distant source would be great, and the resulting flux, even after accounting for the predicted ALP-induced flux enhancement, too low to be detected by CTA under any of the possible array configurations. On the other hand, the attenuation on the EBL is not important below $\sim$100 GeV even for the most distant sources detected to date by IACTs, so the spectra at these low energies probably would not be useful for ALP searches with CTA

As a final remark, it should be noted that since photon/ALP conversion will be used as the main vehicle in this ALP search rather than the usual self-annihilations for WIMP candidates, the unknown dark matter nature will be scrutinized from a different perspective. Indeed, and very interestingly, CTA could probe a region of the ALP parameter space in which ALPs might account for the total cold dark matter content in the Universe.\\

\noindent We acknowledge comments from Seth Digel, Lucy Fortson and Michael Prouza.

\end{document}